\documentclass[prl,preprint,superscriptaddress,showpacs]{revtex4}
\def\beq{\begin{equation}}
\def\eeq{\end{equation}}
\def\beqn{\begin{eqnarray}}
\def\eeqn{\end{eqnarray}}
\usepackage{amsmath}
\usepackage{graphicx}
\usepackage{amssymb}
\usepackage{epsfig}
\usepackage{natbib}
\usepackage[usenames]{color}
\def\mt{$\mathcal{T}$\,}
\def\mi{$\mathcal{I}$\,}

\begin{document}
\title{Berry phase mechanism for optical gyrotropy \\ in stripe-ordered cuprates}
\begin{abstract}
Optical gyrotropy, the lifting of degeneracy between left and right circularly polarized light, can be generated by either time-reversal or chiral symmetry breaking.  In the high-$T_c$ superconductor La$_{2-x}$Ba$_x$CuO$_4$ (LBCO), gyrotropy onsets at the same temperature as charge stripe order, suggesting that the rotation of the stripe direction from one plane to the next generates a helical pattern that breaks chiral symmetry.  In order to further test this chiral stacking hypothesis it is necessary to develop an understanding of the physical mechanism by which chirality generates gyrotropy.  In this paper we show that optical gyrotropy is a consequence of Berry curvature in the momentum space of chiral metals.  We describe a physical picture showing that gyrotropy in chiral metals is closely related to the anomalous Hall effect in itinerant ferromagnets.  We then calculate the magnitude of the gyrotropic response for a given Berry curvature using the semiclassical picture of anomalous velocity and Boltzmann transport theory.  To connect this physical picture with experiment, we calculate the Berry curvature in two tight-binding models. The first model is motivated by the structure of LBCO and illustrates how the gyrotropy is created when the stripe perturbations are added to a simple cubic model.  In the second model, we examine the dramatic enhancement of the gyrotropic coefficient when Rashba spin-orbit coupling is introduced.  The magnitude of the rotation of polarization on reflection expected based these models is calculated and compared with experimental data.
\end{abstract}

\author{J. Orenstein}
\affiliation{Department of Physics, University of California, Berkeley, Berkeley, CA 94720}
\affiliation{Materials Sciences Division, Lawrence Berkeley National Laboratory, Berkeley, CA 94720}
\author{Joel E. Moore}
\affiliation{Department of Physics, University of California, Berkeley, Berkeley, CA 94720}
\affiliation{Materials Sciences Division, Lawrence Berkeley National Laboratory, Berkeley, CA 94720}

\maketitle

\subsection{I. Significance of optical gyrotropy in cuprate superconductors}

Optical gyrotropy is the breaking of degeneracy between left and right circularly polarized light in media, giving rise to phenomena such as the rotation of polarization with propagation \cite{LL} (Faraday effect and optical activity) and upon reflection \cite{Zheludev} (Kerr effect).  Such gyrotropic effects are among the most sensitive and unambiguous probes of symmetry breaking in condensed matter systems.  Currently, measurements of the Kerr effect in cuprate superconductors \cite{Kerr_review} are playing a central role in the effort to understand whether the properties associated with the mysterious pseudogap derive from some form of spontaneous symmetry breaking. The other time-honored tests for symmetry-breaking phase transitions, for example sharp structure in the temperature dependence of the resistivity, static magnetic susceptibility and specific heat, have, for the most part, found negative results despite intensive searches. On the other hand, scattering experiments, with neutrons \cite{Fauque,Y_Li} and X-rays \cite{Ghiringhelli,Chang,Achkar} as probes, find that magnetic and charge-density wave correlations grow with decreasing temperature, hinting at incipient symmetry breaking. However, the finite energy resolution of scattering experiments places limits on identifying truly static order, i.e., the correlation length observed with energy resolution $\Delta E$ is insensitive to fluctuations on time scales that are long compared with $\hbar/\Delta E$.

Coherent optical measurements such as Kerr rotation are capable of discerning order with a time resolution limited only by the patience of the experimenter.  Unlike scattering experiments that are based on particle counting, the Kerr effect is sensitive to the sign of the order parameter.  Thus, signal from fluctuating order will be diminished by the factor $(\tau_\phi/\tau_{exp})^{1/2}$, where $\tau_\phi$ and $\tau_{exp}$ are the experimental averaging time and the order parameter correlation time, respectively.  An experimental averaging time of one second rejects fluctuations on the scale of $\hbar/(1 meV)$ by a factor of about $10^6$.

Kapitulnik and collaborators have reported the onset of Kerr rotation at a temperature, $T_K$, in a variety of underdoped cuprates \cite{Kerr_review,Kerr_YBCO,Kerr_LBCO}, indicating the appearance of truly static long-range order.  However, the question of which symmetry is broken at $T_K$ remains an open one, as two fundamentally different types of symmetry breaking - time-reversal and chiral - can generate optical gyrotropy. Time-reversal breaking in condensed matter is associated with some form of magnetism, whereas chirality corresponds to a loss of mirror symmetries and development of "handedness."  In transparent media these two possible origins can be distinguished by performing a pair of time-conjugate experiments, for example comparing the polarization state of beams of light that propagate through a sample in opposite directions.  Unfortunately, the cuprate superconductors are highly opaque in the near-infrared regime, and optical gyrotropy has been observed only in the reflection, or Kerr effect, geometry.  However, gyrotropy can still be linked with time-reversal if the sign of the rotation depends on the direction of a magnetic field applied as a sample is cooled through the symmetry breaking temperature.

In all the systems studied thus far, cooling in a magnetic field applied just above $T_K$ does not affect the sign of the Kerr rotation \cite{Kerr_review}. This observation tends to rule out the simplest possible interpretation of the gyrotropic response, namely that ferromagnetic order appears at $T_K$. However, it was reported that in YBa$_2$Cu$_3$O$_{7-\delta}$ the sign of the rotation angle, $\theta_K$, reversed when the field was applied at a temperature well above $T_K$, indicating that some form of magnetic order might be involved \cite{Kerr_YBCO}. This observation suggests that the Kerr phenomena could be related to the other piece of evidence for time-reversal breaking in the cuprates, namely the observation of antiferromagnetic (AF) alignment within the unit cell through measurements of spin-flip neutron scattering (SFNS) \cite{Y_Li,Fauque}.  It was pointed out recently that certain forms of AF order consistent with SFNS also break inversion symmetry, giving rise to a magnetoelectric medium in which the Kerr effect is allowed \cite{Kerr_magnetoelectric}. Such a state would be be insensitive to applied magnetic fields acting alone, but could be "trained" by magnetic fields acting together with another perturbation, such as a surface electric field, that couples to inversion breaking order.

Recently reported measurements \cite{Kerr_LBCO} on La$_{2-x}$Ba$_x$CuO$_4$ (LBCO) suggest that this material may provide a new perspective for the interpretation of Kerr phenomena in the cuprates.  In contrast with most members of the cuprate family, LBCO undergoes a series of phase transitions with clearly resolved signatures in transport, thermodynamic, and scattering probes \cite{LBCO_stripes1,LBCO_stripes2}.  Upon cooling LBCO exhibits a stripe-like charge-density-wave state at $T_{CO}$, a spin-ordered state at $T_{SO} < T_{CO}$, and finally a superconductor at $T_{SC} < T_{SO}$.  The striking observation is that $T_K$ coincides with the breaking, at $T_{CO}$, of spatial, rather than time-reversal, symmetry. This observation, taken together with the insensitivity to magnetic field, has led to the suggestion that the observed $T_K$ may reflect broken chiral rather than time-reversal symmetry \cite{Stanford_paper}. A natural source of handedness would be helical stacking of charge density waves (or stripes) that form in each CuO$_2$ layer. However, if the stripes preserve the mirror symmetries of the plane, the symmetry of each layer can be represented by a double-headed arrow and the stacking of such arrows with 90 degree rotation does not form a chiral structure.  Breaking chiral symmetry in LBCO requires additionally that stripes lower the symmetry of the plane, not only by breaking 4-fold rotation, but by removing one of the mirror planes as well.

Although symmetry arguments dictate that a gyrotropic response is allowed in the chiral structure described above, they offer no insight as to the size of $\theta_K$, nor its dependence on any parameters of the electronic structure, such as the amplitude of the charge density modulation.  In order to test the chiral stacking hypothesis by experiment it is necessary to develop an understanding of the underlying physical mechanism for the optical response. In this paper we show that \textit{optical gyrotropy is a consequence of the Berry curvature in the momentum space of chiral metals}.  In Section II we describe a physical picture that links gyrotropy in chiral metals with the anomalous Hall effect in itinerant ferromagnets.  We then calculate the magnitude of the gyrotropic response for a given Berry curvature using the semiclassical picture of anomalous velocity and Boltzmann transport theory.  In Section III we present a calculation of the Berry curvature in two tight-binding models.  The first model is motivated by the physics of LBCO and has no spin-orbit coupling; the point is to show how the gyrotropic effect is created when the stripe perturbations are added to a simple cubic model. In the second model, we investigate the effect of including Rashba spin-orbit coupling on the Berry curvature.  Finally, in Section IV we estimate the magnitude of the Kerr rotation as a function of optical frequency expected based on the calculated Berry curvature and compare with experimental data.

\subsection{II. Description of Berry mechanism for optical gyrotropy}

The theory of electron transport based on the Fermi liquid picture of Landau was recently understood to be an incomplete description, as it did not include the anomalous velocity \cite{Karplus,Sundaram,Panati,Teufel} associated with Berry curvature \cite{Berry}.  It is now generally accepted that the velocity of an electron wavepacket in a band with dispersion $\varepsilon({\bf k})$ is given by,

\beq
{\bf v}({\bf k})={1 \over \hbar}{\partial \varepsilon({\bf k})\over \partial {\bf k}}-{e\over \hbar}\bf {E}\times\bf{\Omega}(\bf{k}).
\label{anomvel}
\eeq

The extra, "anomalous velocity" term is the cross product of the electric field, $\bf{E}$, with the Berry curvature, $\bf {\Omega}(\bf{k})$. The anomalous velocity originates from the variation with wavevector of the electron's wavefunction within each unit cell, that is, the dependence of the Wannier orbitals on $\bf{k}$.

The Berry curvature is a time-odd, axial vector, transforming under time reversal (\mt) and inversion (\mi) according to \cite{Niu_Berry_review},
\beq
\hbox{Under \mt}: \bf{\Omega}(\bf{k})\rightarrow -\bf{\Omega}(\bf{-k})
\label{Omega_symmetry}
\eeq
\beq
\hbox{Under \mi}: \bf{\Omega}(\bf{k})\rightarrow \bf{\Omega}(\bf{-k}).
\label{Omega_symmetry}
\eeq
Such transformation properties imply that $\bf {\Omega}(\bf{k})$ vanishes in systems that respect both symmetries. In metals that break \mt but preserve \mi the flux of $\bf {\Omega}(\bf{k})$ through a contour of constant energy will be nonzero.  This nonzero flux is responsible for the intrinsic anomalous Hall effect (AHE) in magnetic metals \cite{Sinova_AHE_review}, where the off-diagonal conductivity is given by,

\beq
\sigma_{xy}^{AHE}={e^2 \over \hbar}\int d^3k\,\Omega_z(\textbf{k})f(\varepsilon_{\textbf{k}}),
\label{@D_AHE}
\eeq
and $f(\varepsilon_{\bf{k}})$ is the Fermi occupation function.

Lesser known, but potentially equally significant, are phenomena associated with metals that preserve \mt symmetry but break \mi. Figure 1 shows a sketch of the symmetry-allowed Berry curvature on the Fermi surface (FS) of a two-dimensional electron gas that breaks \mi but preserves \mt. Because $\bf {\Omega}(\bf{k})$ is an odd function of $\bf{k}$, the anomalous velocity generated by a uniform electric field is also odd and therefore the linear response (AHE) vanishes.  However, it was shown recently \cite{Moore_CPGE} that the Berry curvature and associated anomalous velocity provide an intrinsic mechanism for the nonlinear "photogalvanic effects" in metals \cite{CPGE_review}.  In these effects, excitation with an oscillating electric field generates a dc photocurrent in the absence of an applied voltage that is proportional to the field amplitude squared and whose direction depends on its polarization state.  The photogalvanic current appears at second order in $E$, as one power of $E$ acts to shift the FS and the AHE of this shifted FS is non-zero.

\begin{figure}
\includegraphics[width=3.0in]{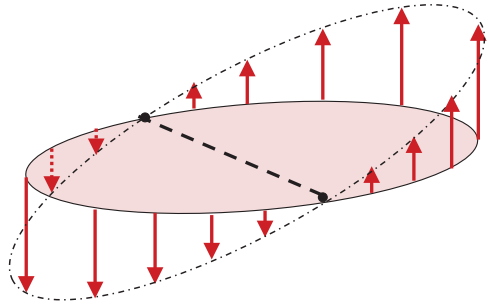}
\caption{Filled momentum space eigenstates of a two-dimensional metal shown as the shaded region. Red arrows indicate the Berry curvature on the Fermi contour that marks the boundary of occupied states.  For a system that preserves time-reversal, but breaks inversion symmetry, the Berry curvature is an odd function of the electron wavevector.}
\end{figure}
Below we show that, in addition to nonlinear photogalvanic currents, the Berry curvature generates a linear in $E$ effect, which is optical gyrotropy.  We can then summarize the linear response effects of metals generated by the Berry curvature as follows:

\beq
\hbox{\mt breaking:}\  j_x=\sigma_{xy}^{AHE}E_y
\label{Constitutive}
\eeq
\beq
\hbox{\mi breaking:}\ j_x=\lambda_{xyz}^{G}{dE_y \over dz}.
\label{Constitutive}
\eeq
The transverse current that appears in Eq. 6, with linear response coefficient $\lambda_{ijk}^G$, is proportional to the spatial variation of the electric field, that is, it is a non-local effect.  Propagation of light through a medium with nonzero $\lambda_{ijk}^G$ is described by combining the non-local constitutive relation with Maxwell's equations.  For propagation along the optic axis of a uniaxial medium the normal modes of propagation are the two transverse circularly polarized waves, with different indices of refraction obtained for the left and right-handed modes \cite{LL}.  This index difference is sufficient to yield polarization rotation on reflection \cite{Zheludev}.

To calculate $\lambda_{ijk}^G$ we consider cuts at different values of $k_z$ through a 3D Fermi surface.  The flux of $\bf{\Omega}$ through the surface defined by $k_z = 0$ vanishes, as was the case for the isolated 2D layer shown in Fig. 1.  However, this restriction does not apply to the Fermi contours at nonzero $k_z$ because the points at $k_z,\pm \bf{k}$, where $\bf{k}$ is the wavevector in the $xy$ plane, are not related by time-reversal.  As a result, the flux, $\Phi(k_z)$, of the Berry curvature through a Fermi contour with nonzero $k_z$ will in general be nonzero, although it is required by time-reversal to be an odd function of $k_z$.  Thus, with the exception of $k_z=0$, each horizontal slice through the Fermi surface contributes a nonzero anomalous Hall conductivity.  The vanishing of the net Hall current required by \mt symmetry is restored with integration on $k_z$.

\begin{figure}
\includegraphics[width=3.0in]{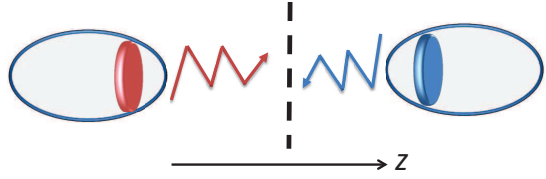}
\caption{Representations in position and momentum space of the origin of transverse non-local current in a chiral metal. Jagged arrows indicate the diffusive motion of electrons in real space.  Ovals represent the 3D Fermi surface.  The red and blue slices in each oval indicate the the regions of momentum space occupied by electrons that reach a given $z$ from $z^\prime<z$ and $z^\prime>z$, respectively.}
\end{figure}

Although the AHE is zero, this system will manifest a nonzero gyrotropic coefficient, as shown below.  To focus on the effects of the electric field gradient, we consider the transport in position space, as illustrated in Fig. 2. We assume an applied electric field of the form $E_y(z,t)=Re\{E\exp [i(qz-\omega t)]\}$ and follow a Boltzmann approach that is valid in when $qv_z \tau\ll 1$, where $v_z$ and $\tau_z$ are the electron's velocity and mean-free time for transport in the $z$-direction, respectively.  Nonlocality arises from the assumption that an electron arriving at $z$ from $z^\prime$ without scattering retains the anomalous velocity induced by $E(z')$.  The crucial point is that electrons arriving from $z^\prime <z$ have positive $v_z$ while electrons that arrive from $z^\prime>z$ have negative $v_z$.  As a result electrons arriving at $z$ from the left and right originate from regions of the FS with opposite signs of $k_z$ and therefore with opposite anomalous velocity.  In the presence of an electric field gradient the net anomalous Hall current at $z$ will not cancel, and thus a nonzero gyrotropic coefficient is obtained.

To calculate the gyrotropic coefficient within this model, we first determine the transverse current carried by electrons whose wavevectors lie in a slice of $k$-space of thickness $dk_z$ assuming local electrodynamics,

\beq
J_{loc}(k_z,z,t)=-{e^2\over\hbar}{dk_z\over 4 \pi^2}E\exp[i(qz-\omega t)]\Phi(k_z),
\label{Local_current}
\eeq
where,
\beq
\Phi(k_z)=\int^{k_F(k_z)}_{0}d^2k\,\Omega_z(k,k_z).
\label{Phi_defn}
\eeq
The transverse nonlocal current, $J_{nl}(z,t)$, that results from electrons accelerated at $z^\prime > z$ is given by,
\beq
J_{nl}(k_z,z,t)=\int_{z}^{\infty} dz^\prime p(z-z^\prime)J_{loc}[k_z, z^\prime, t^\prime(z^\prime)]),
\label{Nonlocal_current}
\eeq
where $t^\prime\equiv t-|(z^\prime-z)/v_z(k_z)|$. The same expression integrated from $z$ to $-\infty$ applies to electrons with $z^\prime <z$. In Eq. 9, $p(z-z^\prime)$ is the probability that an electron will arrive at $z$ if its last scattering event was at $z^\prime$.  Within the relaxation time approximation $p(z-z^\prime)\propto e^{-|z-z^\prime|/|v_z \tau|}$.  Substituting this form of $p(z-z^\prime)$ into Eq. 9 and summing the contributions from the left and right moving electrons yields,

\beq
J_{nl}(k_z,z,t)=-{e^2\over\hbar}{dk_z\over 4 \pi^2}\Phi(k_z){2iqv_z(k_z)\tau_z\over(1-i\omega\tau_z)^2}E\exp[i(qz-\omega t)],
\label{Nonlocal_slice}
\eeq
in the $qv_z(k_z)\tau\ll1$ limit. The net current is obtained by integrating the above over $k_z$, yielding a nonlocal response coefficient,

\beq
\lambda^G_{xyz}={e^2\over h}{1\over(1-i\omega\tau_z)^2}\int^{k_{Fz}}_{-k_{Fz}}dk_z\Phi(k_z)v_z(k_z)\tau_z.
\label{Nonlocal_slice}
\eeq

This result for the gyrotropic effect induced by Berry curvature is distinct from that recently found by Mineev and Yoshioka~\cite{mineev} in a Kubo formula calculation of the optical conductivity of metals with linear-in-wavevector spin-orbit coupling, $\bf\sigma\cdot\gamma_0 \bf{k}$, which is allowed in non-centrosymmetric media.  They obtained a gyrotropic coefficient in the proportional to $\omega/\gamma_0 k_F$, where $k_F$ is the Fermi wavevector, and with no appearance of the scattering time $\tau$.  While the Berry curvature is not explicit in that calculation, it is conceivable that the result obtained originates from the Berry curvature, but in a different high-frequency limit ($\omega \tau \gg 1$) in which $\tau$ is unimportant.  We proceed to understand the magnitude of the gyrotropic effect in two models of inversion symmetry breaking.

\subsection{III. Model calculations}
\subsubsection{III.A Introduction}

In a layered material, some intuition for the gyrotropic effect can be obtained by regarding the interplane coupling as a perturbation to the two-dimensional problem of a single plane.  Since time-reversal symmetry is unbroken, each plane has zero anomalous Hall effect: if inversion symmetry is broken, there will be points of nonzero Berry curvature, but under these conditions $\bf \Omega({\bf k}) = -\bf \Omega(-{\bf k})$ while $\varepsilon({\bf k}) = \varepsilon(-{\bf k})$, guaranteeing that the integral of $\Omega$ over occupied states vanishes.  Turning on the interplane coupling, at generic values of $k_z$ time-reversal is broken, which means the $z$-coupling can be regarded as a \mt breaking perturbation to the 2D band structure.

For a large gyrotropic effect, the perturbation should induce a large (of order $e^2/h$) anomalous Hall effect in most $k_z$ planes viewed as 2D band structures, that is, the decoupled 2D plane must be highly susceptible to some time-reversal-breaking perturbation that yields an AHE.  This motivates the model of coupled Rashba 2DEGs \cite{Rashba} described below, in which the interplane coupling acts like a Zeeman magnetic field, which is well known to produce a large AHE \cite{Dugaev_AHE,Culcer_AHE}.   Spin-orbit coupling in this model provides a large magnitude of the Berry curvature as the wavefunction evolves rapidly with crystal momentum; we will see that this model realizes the full natural magnitude of the gyrotropic effect, analogous to $\sigma_{xy} = e^2/h$ for the Hall effect. Conversely, we expect a relatively small gyrotropic effect if the 2D planes are not particularly close to any state with a large AHE.  In the model below of stripe order in a cubic lattice, inspired by the cuprates, we indeed obtain a nonzero but small gyrotropic effect in the lowest band near the $\Gamma$-point.

\subsubsection{III.B Spinless model}

The first tight-binding model we consider has no spin-orbit coupling and spin indices are suppressed in the following.  The starting point is a cubic lattice of spacing $a$ with a single orbital per site and nearest-neighbor hopping $t$,
\begin{equation}
H_0 = -t \sum_{\langle ij \rangle} (c_i^\dagger c_j + c_j^\dagger c_i),
\end{equation}
where $i$ and $j$ are the sites of a nearest-neighbor bond.  The resulting band has energy
\begin{equation}
E_0({\bf k}) = -2 t (\cos k_x a + \cos k_y a + \cos k_z a),
\end{equation}
with a minimum at the ${\bf k}=0$ ($\Gamma$) point. If the Fermi energy is slightly above $-6 t$, the Fermi surface is a sphere around the $\Gamma$ point.

\begin{figure}
\includegraphics[width=3.5in]{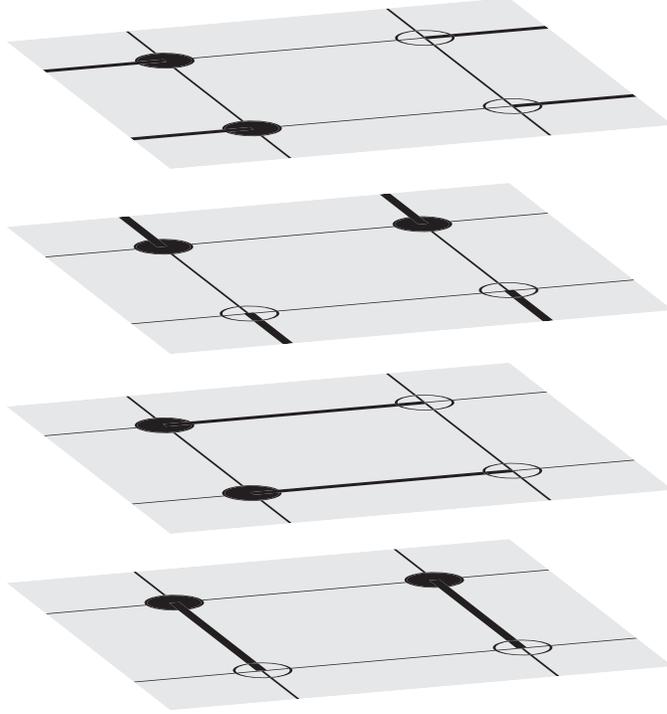}
\caption{Inversion-breaking structure with 16-site unit cell on a cubic lattice. The thick bonds are those with modified hopping $t+\delta t$ compared to $t$ for the thin bonds, and the solid sites have potential $\epsilon$ compared to zero for the thin bonds.  Each individual plane has no inversion center if and only if $\epsilon$ and $\delta t$ are both nonzero, and under these conditions the structure has a nonzero a gyrotropic effect.}
\label{bandfig}
\end{figure}

The key quantity for rotation of the plane of polarization for light incident along ${\bf \hat z}$ is the $z$-component of the Berry curvature,
\begin{equation}
\Omega_z({\bf k}) = -i \left(\langle \partial_{k_x} u  | \partial_{k_y} u \rangle -  \langle \partial_{k_y} u | \partial_{k_x} u \rangle \right),
\end{equation}
where $u$ is the periodic part of a Bloch state $\psi_{\bf k} = \exp(i {\bf k} \cdot {\bf r}) u_{\bf k}({\bf r})$.  The Berry curvature is identically zero in a model with both inversion and time-reversal symmetry, such as the unperturbed $H_0$; we seek to compute the $\Omega_z$ generated by symmetry-lowering perturbations.  Consider now a model where the unit cell is increased to 16 sites and the bonds indicated by thick lines in Fig.~\ref{bandfig} have hopping matrix element $- (t + \delta t)$ rather than strength $-t$.  We also add an additional on-site potential $\epsilon$ on the circled sites.  Labeling the set of thick bonds in Fig.~\ref{bandfig} by $B$, the sites of a bond $b$ by $b1$ and $b2$, and the set of solid-circle sites $S$, we have the full Hamiltonian,
\begin{equation}
H = H_0 - (\delta t) \sum_{b \in B} (c_{b1}^\dagger c_{b2} + c_{b2}^\dagger c_{b1}) + \epsilon \sum_{s \in S} c^\dagger_s c_s.
\end{equation}
Each $xy$ plane has a single mirror if both $\delta t$ and $\epsilon$ are nonzero, and there is a screw axis parallel to ${\bf z}$ and passing through the upper left sites in Fig.~\ref{bandfig}.  If either of the perturbations is zero, then there is an inversion center and $\Omega_z$ vanishes.

There are straightforward procedures to calculate $\Omega_z$ anywhere in the Brillouin zone: the simplest is to find smooth wavefunctions $u$ in a sufficiently large patch of the Brillouin zone to evaluate $\Omega_z$, and a robust alternative is to use the projection operator onto the Bloch eigenstate for an explicitly gauge-invariant calculation~\cite{avronseilersimon}.  To get a simple analytical result, let us stick to the vicinity of the $\Gamma$ point and treat the bond- and site-modulations perturbatively.  A smooth gauge is obtained in that region if we require that the first component of the Bloch eigenvector in the site basis be positive.  The result of the $\Omega_z$ calculation, which requires computing the wavefunctions to order $k_x k_y k_z$, is that the leading behavior for small ${\bf k} a$, $(\delta t)/t$, and $\epsilon/t$ is
\begin{equation}
\Omega_z ({\bf k}) = {289 \over 3456000} \left({\delta t \over t}\right)^2 \left({\epsilon \over t}\right)^2 a^3 k_z.
\label{smallcoeff}
\end{equation}
This is consistent with the expectation that the leading-order gyrotropic effect should not depend on the sign of either perturbation (as the direction of the screw axis does not change) and should require both perturbations to be present.

The smallness of the dimensionless prefactor in this model (approximately $8.36 \times 10^{-5}$) indicates that the $k_z \not = 0$ planes are quite far removed from quantum anomalous Hall 2D band structures, which would have integer Chern number (the Brillouin zone integral of $\Omega_z$ divided by $2 \pi$).  We have also computed a similar model where the screw axis passes through faces rather than sites; in this model the coefficient in Equation (\ref{smallcoeff}) is increased by $(27/17)^2$.  This suggests that the strength of the gyrotropic effect could be increased in materials where the Fermi level is closer to a band crossing, as significant Berry curvature accumulates in the vicinity of degeneracies or avoided crossings.  Incommensurate electronic stripes are an interesting limit where inversion is infinitesimally broken in the sense that each plane has a point that is arbitrarily close to being an inversion center, so the gyrotropic effect will vanish.  Locking of the stripes will then be accompanied by an increase in the gyrotropic effect if the locked pattern breaks inversion.

\subsubsection{III.C Tight-binding model including spin-dependent hopping}

We now consider a 2D square-lattice tight-binding model with an additional spin-dependent hopping term that generates Rashba spin-orbit coupling:
\begin{equation}
H = -t \sum_{\langle ij \rangle,\sigma} (c_{i\sigma}^\dagger c_{j \sigma} + c_{j \sigma}^\dagger c_{i \sigma}) + t_R \sum_j \left(i c^\dagger_{j \sigma} s^y_{\sigma \sigma^\prime} c_{(j+{\bf \hat x}) \sigma^\prime} + {\rm h.c.}\right) - \left(i c^\dagger_{j \sigma} s^x_{\sigma \sigma^\prime} c_{(j+{ \bf \hat y}) \sigma^\prime} + {\rm h.c.}\right).
\end{equation}
To this we add a spin-dependent interplane coupling
\begin{equation}
H_z =  t_z \sum_j \left(i c^\dagger_{j \sigma} s^z_{\sigma \sigma^\prime} c_{(j+{\bf \hat z}) \sigma^\prime} + {\rm h.c.}\right).
\end{equation}
Writing $\lambda_R = t_R \hbar / 2$, $\lambda_z = t_z \hbar / 2$, the $2\times2$ Bloch Hamiltonian for $H+H_z$ takes a simple form in terms of the Pauli matrices:
\begin{equation}
H_B = -2 t \left[\cos(k_x a) + \cos(k_y a)\right] {\bf 1} + \lambda_R \left[\sin (k_x a) \sigma_y -  \sin (k_y a) \sigma_x \right] + \lambda_z \sin(k_z a) \sigma_z.
\label{rashbaeq}
\end{equation}

\begin{figure}
\includegraphics[width=6.5in]{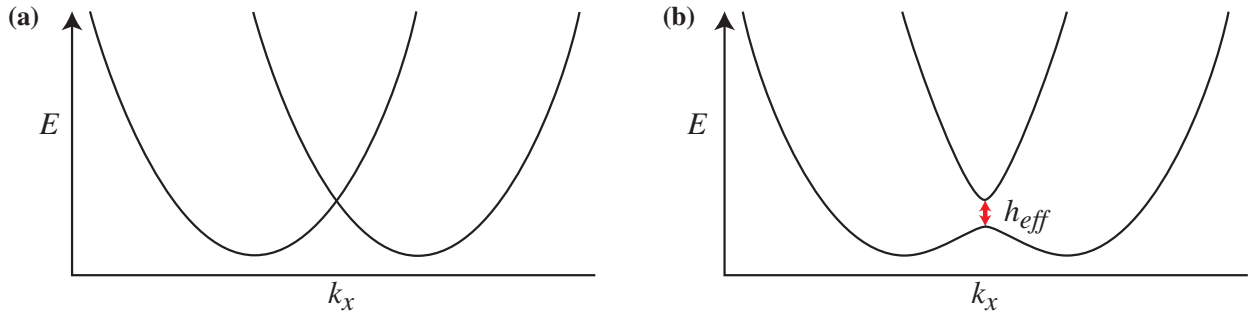}
\caption{(a) Slice of the two-dimensional band structure described by a Hamiltonian with quadratic and Rashba terms.  (b) The effect of a $z$-directed Zeeman field, which opens a gap at the $\Gamma$ point proportional to the field strength.  This is equivalent to the effect of the $z$-directed spin-orbit-coupled hopping in the model described by equation (\ref{rashbaeq}), which can be viewed as an effective field $h_{\rm eff}(k_z)$.}
\label{figrashba}
\end{figure}

We will study the behavior of this band structure when $\lambda_z \ll t, \lambda_R$ and when $k_x a$ and $k_y a$ are both much less than unity.  At fixed $k_z$ this becomes exactly the problem of a Rashba 2DEG in a $z$-directed Zeeman field, and a standard result from that literature will imply immediately that this model can have a strong gyrotropic effect.  The $z$-directed Zeeman field modifies the band structure near the $\Gamma$ point as shown in Fig.~\ref{figrashba}.  Writing ${\tilde E} = E_F + 2 t$ for the energy above the band minimum and setting $a=1$, we find a quadratic equation for $R^2 = k_x^2 + k_z^2$ on the Fermi surface, with solutions
\begin{equation}
\label{quadsoln}
R^2 = {2 t {\tilde E} + {\lambda_R}^2 \pm \sqrt{{\lambda_R}^4 + 4  {\lambda_R}^2 t {\tilde E}  +4 {\lambda_z}^2 \sin^2 k_z t^2} \over 2 t^2}.
\end{equation}
(When one or both values on the right-hand-side are negative, the Fermi surface contains one or zero sheets respectively.)
We see that the 2D $\Gamma$ point $R=0$ always lies on the Fermi surface at ${\tilde E}=0$ if $\sin k_z = 0$.  At this Fermi energy the physics is quite simple: when $\sin k_z$ is positive, the 2D band structure has $\sigma_{xy} = e^2 / (2 h)$, and when $\sin k_z$ is negative, $\sigma_{xy} = - e^2 /h$.

These results follow from computing that the AHE strength for the 2D band structure at constant $k_z$ is
\begin{equation}
\sigma_{xy} = {e^2 \over h} {\theta(R_1) - \theta(R_2) \over \pi},
\label{anglediff}
\end{equation}
where ${R_1}^2$ and ${R_2}^2$ are the larger and smaller nonnegative solutions of (\ref{quadsoln}): ($R_2 = 0$ if only one nonnegative solution exists), and
\begin{equation}
\theta(R) = \arctan\left({\lambda_R R \over \lambda_z \sin k_z}\right).
\end{equation}
The geometrical picture of this result is that the integral of $\Omega_z$ over a partially filled 2D band is proportional to the area on the Bloch sphere swept out by the corresponding 2-component eigenspinor.  For our rotationally symmetric case, that area is just determined by the highest and lowest lines of latitude reached, which gives (\ref{anglediff}).

In the following section we restore the lattice spacing and dielectric constant and convert this result to a Kerr rotation measurable in experiment.  The fact that a small perturbation $\lambda_z$ can lead to a strong gyrotropic effect in this model if ${\tilde E}=0$ is a consequence of the degeneracy of the unperturbed band structure; the smallness of the effect in the previous example essentially results because the starting band structure (the cubic tight-binding model near the $\Gamma$ point) is quite stable.  We conclude from this example of weakly coupled Rashba 2DEGs that the fundamental scale of the Berry phase entering the gyrotropic effect, when the symmetry breaking is fully developed, corresponds to $\Omega_z$ of order $a^2$ in a generic constant-$k_z$ plane of the Brillouin zone.
\subsection{IV. Estimating the size of the Kerr rotation}

The gyrotropic response of chiral media is usually expressed by a non-local dielectric tensor $\gamma_{ijk}$ defined in terms of an expansion of the electric displacement in orders of the spatial derivative of the electric field,
\begin{equation}
D_i=\epsilon_{ij}E_j+\gamma_{ijk}{dE_j\over dx_k}+...
\end{equation}
According to Ref. \cite{Zheludev}, the Kerr effect at normal incidence on the optic axis of a uniaxial crystal is given by,
\begin{equation}
\theta_K={\omega \over c} Im \left\{{\gamma_{ijk} \over \epsilon_\|(\omega)-1}\right\},
\end{equation}
where $\epsilon_\|(\omega)$ is the dielectric function in the plane perpendicular to the optic axis.  To estimate the maximum values of $\theta_K$ that can be expected, we use the result of the previous section, in which it was shown that
\begin{equation}
\int^{k_{Fz}}_{-k_{Fz}}dk_z \Phi(k_z)\sim 1,
\end{equation}
when the Fermi level lies between two bands that are split by a nonzero $k_z$.  Substituting this value into Eq. 9 yields,
\begin{equation}
\theta_K(\omega) \sim \alpha l_z Re \left[{1 \over (1-i\omega\tau_z)^2[\epsilon_\|(\omega)-1]}\right],
\end{equation}
where $\alpha$ is the fine structure constant and $l_z$ is the mean-free-path in the $z$-direction, normalized to the lattice constant.

\begin{figure}
\includegraphics[width=3.0in]{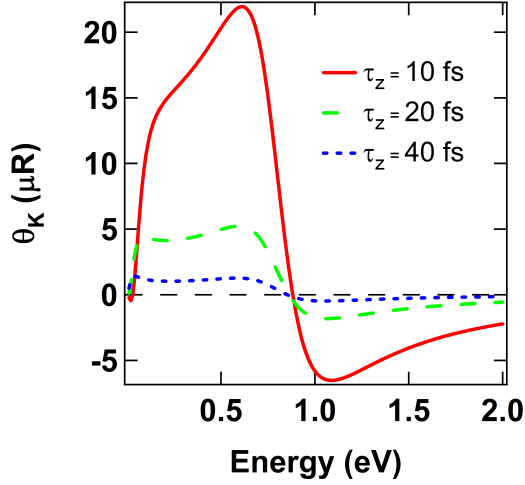}
\caption{Spectrum of $\theta_K(\omega)$ based on Eq. 27, for the three values of $\tau_z$ indicated in the legend. The zero crossing in the vicinity of the plasma frequency is a consequence of the proportionality of $\theta_K$ to the real part of $[\epsilon_\|(\omega)-1]^{-1}$ in this frequency range. The photon energy of 0.8 eV used in the Kerr experiments lies just below the energy of the predicted zero-crossing.}
\end{figure}

Fig. 5 shows the spectrum of $\theta_K$ expected on the basis of Eq. 26, with transport parameters chosen to test for consistency with Kerr data on LBCO. As the transport in the direction perpendicular to the CuO$_2$ planes is incoherent, we set $l_z=1$. We take $\hbar/t_\bot$ to be a lower bound on $\tau_z$; a typical value for the interplane hopping matrix element, $t_\bot =$ 70 meV corresponds to $\tau_z=0.01$ ps. The spectrum of the in-plane dielectric function, $\epsilon_\|(\omega)$ is obtained from Ref. \cite{Homes}. Both the strong dependence of the Kerr rotation on $\tau_z$ and the zero crossing in the spectrum of $\theta_K(\omega)$ are expected in the high frequency ($\omega\tau_z \gg 1$) limit of Eq. 26, where,

\begin{equation}
\theta_K(\omega) \rightarrow -{\alpha l_z \over (\omega\tau_z)^2}Re \left\{{1 \over(\epsilon_\|(\omega)-1)}\right\}.
\label{angle}
\end{equation}

The experimentally determined Kerr rotation in LBCO, $\theta_K=6$ $\mu$rad at $\hbar \omega$=0.8 eV, corresponds essentially to the maximum value possible with our theory, that is the smallest reasonable value of $\tau_z$, coupled with a Berry flux of order unity through generic planes of constant $k_z$. As the Rashba model in which this result obtains is not directly applicable to the cuprate bandstructure, the large value of the Kerr effect obtained experimentally would appear to require some other topological property of the cuprate bandstructure that leads to large Berry curvature. Evidence for the latter would be sensitivity to \mt-breaking perturbations such as magnetic field, and indeed, such evidence can be found in the highly singular nature of the Hall coefficient in the neighborhood of the charge order transition in LBCO \cite{Hall}.  Regarding future experiments, our theory predicts that $\gamma$ is real in the near infrared and, therefore, that $\theta_K(\omega)$ will be proportional to the real part of ${1/[\epsilon_\|(\omega-1)]}$.  Experiments that probe the spectrum of $\theta_K$ in the region near the plasma resonance of the cuprates would be valuable in determining the origin of their gyrotropic response.

The authors acknowledge useful conversations with Pavan Hosur, Aharon Kapitulnik, Steven Kivelson, Patrick Lee, and Srinivas Raghu, and support from the Office of Basic Energy Sciences, Materials Sciences and Engineering Division of the U.S. Department of Energy under Contract No. DE-AC02-05CH11231.


\begin{thebibliography}{30}
\expandafter\ifx\csname natexlab\endcsname\relax\def\natexlab#1{#1}\fi
\expandafter\ifx\csname bibnamefont\endcsname\relax
  \def\bibnamefont#1{#1}\fi
\expandafter\ifx\csname bibfnamefont\endcsname\relax
  \def\bibfnamefont#1{#1}\fi
\expandafter\ifx\csname citenamefont\endcsname\relax
  \def\citenamefont#1{#1}\fi
\expandafter\ifx\csname url\endcsname\relax
  \def\url#1{\texttt{#1}}\fi
\expandafter\ifx\csname urlprefix\endcsname\relax\def\urlprefix{URL }\fi
\providecommand{\bibinfo}[2]{#2}
\providecommand{\eprint}[2][]{\url{#2}}

\bibitem[{\citenamefont{Landau et~al.}(1984)\citenamefont{Landau, Lifshitz, and
  Pitaevskii}}]{LL}
\bibinfo{author}{\bibfnamefont{L.}~\bibnamefont{Landau}},
  \bibinfo{author}{\bibfnamefont{E.}~\bibnamefont{Lifshitz}}, \bibnamefont{and}
  \bibinfo{author}{\bibfnamefont{L.}~\bibnamefont{Pitaevskii}},
  \emph{\bibinfo{title}{Electrodynamics of Continuous Media}}
  (\bibinfo{publisher}{Pergamon Press}, \bibinfo{address}{New York},
  \bibinfo{year}{1984}).

\bibitem[{\citenamefont{Bungay et~al.}(1993)\citenamefont{Bungay, Svirko, and
  Zheludev}}]{Zheludev}
\bibinfo{author}{\bibfnamefont{A.~R.} \bibnamefont{Bungay}},
  \bibinfo{author}{\bibfnamefont{Y.~P.} \bibnamefont{Svirko}},
  \bibnamefont{and} \bibinfo{author}{\bibfnamefont{N.~I.}
  \bibnamefont{Zheludev}}, \bibinfo{journal}{Phys. Rev. B}
  \textbf{\bibinfo{volume}{47}}, \bibinfo{pages}{11730} (\bibinfo{year}{1993}).

\bibitem[{\citenamefont{Kapitulnik et~al.}(2009)}]{Kerr_review}
\bibinfo{author}{\bibfnamefont{A.}~\bibnamefont{Kapitulnik}}
  \bibnamefont{et~al.}, \bibinfo{journal}{New. J. Phys.}
  \textbf{\bibinfo{volume}{11}}, \bibinfo{pages}{055060}
  (\bibinfo{year}{2009}).

\bibitem[{\citenamefont{Fauque et~al.}(2006)}]{Fauque}
\bibinfo{author}{\bibfnamefont{B.}~\bibnamefont{Fauque}} \bibnamefont{et~al.},
  \bibinfo{journal}{Phys. Rev. Lett.} \textbf{\bibinfo{volume}{96}},
  \bibinfo{pages}{197001} (\bibinfo{year}{2006}).

\bibitem[{\citenamefont{Li et~al.}(2008)}]{Y_Li}
\bibinfo{author}{\bibfnamefont{Y.}~\bibnamefont{Li}} \bibnamefont{et~al.},
  \bibinfo{journal}{Nature} \textbf{\bibinfo{volume}{455}},
  \bibinfo{pages}{372} (\bibinfo{year}{2008}).

\bibitem[{\citenamefont{Ghiringhelli et~al.}(2012)}]{Ghiringhelli}
\bibinfo{author}{\bibfnamefont{G.}~\bibnamefont{Ghiringhelli}}
  \bibnamefont{et~al.}, \bibinfo{journal}{Science}
  \textbf{\bibinfo{volume}{337}}, \bibinfo{pages}{821} (\bibinfo{year}{2012}).

\bibitem[{\citenamefont{Chang et~al.}(2012)}]{Chang}
\bibinfo{author}{\bibfnamefont{J.}~\bibnamefont{Chang}} \bibnamefont{et~al.},
  \bibinfo{journal}{arXiv:1206.4333}  (\bibinfo{year}{2012}).

\bibitem[{\citenamefont{Achkar et~al.}(2012)}]{Achkar}
\bibinfo{author}{\bibfnamefont{A.~J.} \bibnamefont{Achkar}}
  \bibnamefont{et~al.}, \bibinfo{journal}{Phys. Rev. Lett.}
  \textbf{\bibinfo{volume}{109}}, \bibinfo{pages}{167001}
  (\bibinfo{year}{2012}).

\bibitem[{\citenamefont{Xia et~al.}(2008)}]{Kerr_YBCO}
\bibinfo{author}{\bibfnamefont{J.}~\bibnamefont{Xia}} \bibnamefont{et~al.},
  \bibinfo{journal}{Phys. Rev. Lett.} \textbf{\bibinfo{volume}{100}},
  \bibinfo{pages}{127002} (\bibinfo{year}{2008}).

\bibitem[{\citenamefont{Karapetyan et~al.}(2012)}]{Kerr_LBCO}
\bibinfo{author}{\bibfnamefont{H.}~\bibnamefont{Karapetyan}}
  \bibnamefont{et~al.}, \bibinfo{journal}{Phys. Rev. Lett.}
  \textbf{\bibinfo{volume}{109}}, \bibinfo{pages}{147001}
  (\bibinfo{year}{2012}).

\bibitem[{\citenamefont{Orenstein}(2011)}]{Kerr_magnetoelectric}
\bibinfo{author}{\bibfnamefont{J.}~\bibnamefont{Orenstein}},
  \bibinfo{journal}{Phys. Rev. Lett.} \textbf{\bibinfo{volume}{107}},
  \bibinfo{pages}{067002} (\bibinfo{year}{2011}).

\bibitem[{\citenamefont{Tranquada et~al.}(2008)\citenamefont{Tranquada, Gu,
  H\"ucker, Jie, Kang, Klingeler, Li, Tristan, Wen, Xu et~al.}}]{LBCO_stripes1}
\bibinfo{author}{\bibfnamefont{J.~M.} \bibnamefont{Tranquada}},
  \bibinfo{author}{\bibfnamefont{G.~D.} \bibnamefont{Gu}},
  \bibinfo{author}{\bibfnamefont{M.}~\bibnamefont{H\"ucker}},
  \bibinfo{author}{\bibfnamefont{Q.}~\bibnamefont{Jie}},
  \bibinfo{author}{\bibfnamefont{H.-J.} \bibnamefont{Kang}},
  \bibinfo{author}{\bibfnamefont{R.}~\bibnamefont{Klingeler}},
  \bibinfo{author}{\bibfnamefont{Q.}~\bibnamefont{Li}},
  \bibinfo{author}{\bibfnamefont{N.}~\bibnamefont{Tristan}},
  \bibinfo{author}{\bibfnamefont{J.~S.} \bibnamefont{Wen}},
  \bibinfo{author}{\bibfnamefont{G.~Y.} \bibnamefont{Xu}},
  \bibnamefont{et~al.}, \bibinfo{journal}{Phys. Rev. B}
  \textbf{\bibinfo{volume}{78}}, \bibinfo{pages}{174529}
  (\bibinfo{year}{2008}).

\bibitem[{\citenamefont{M.~H\"ucker et~al.}(2011)\citenamefont{M.~H\"ucker,
  v.~Zimmermann, Gu, Xu, Wen, Xu, Kang, Zheludev, and
  Tranquada}}]{LBCO_stripes2}
\bibinfo{author}{\bibfnamefont{M.}~\bibnamefont{M.~H\"ucker}},
  \bibinfo{author}{\bibfnamefont{M.}~\bibnamefont{v.~Zimmermann}},
  \bibinfo{author}{\bibfnamefont{G.~D.} \bibnamefont{Gu}},
  \bibinfo{author}{\bibfnamefont{Z.~J.} \bibnamefont{Xu}},
  \bibinfo{author}{\bibfnamefont{J.~S.} \bibnamefont{Wen}},
  \bibinfo{author}{\bibfnamefont{G.}~\bibnamefont{Xu}},
  \bibinfo{author}{\bibfnamefont{H.~J.} \bibnamefont{Kang}},
  \bibinfo{author}{\bibfnamefont{A.}~\bibnamefont{Zheludev}}, \bibnamefont{and}
  \bibinfo{author}{\bibfnamefont{J.~M.} \bibnamefont{Tranquada}},
  \bibinfo{journal}{Phys. Rev. B} \textbf{\bibinfo{volume}{83}},
  \bibinfo{pages}{104506} (\bibinfo{year}{2011}).

\bibitem[{\citenamefont{Hosur et~al.}(2012)}]{Stanford_paper}
\bibinfo{author}{\bibfnamefont{P.}~\bibnamefont{Hosur}} \bibnamefont{et~al.}
  (\bibinfo{year}{2012}), \eprint{arXiv:1212.2274}.

\bibitem[{\citenamefont{Karplus and Luttinger}(1954)}]{Karplus}
\bibinfo{author}{\bibfnamefont{R.}~\bibnamefont{Karplus}} \bibnamefont{and}
  \bibinfo{author}{\bibfnamefont{J.~M.} \bibnamefont{Luttinger}},
  \bibinfo{journal}{Phys. Rev. B} \textbf{\bibinfo{volume}{95}},
  \bibinfo{pages}{1154} (\bibinfo{year}{1954}).

\bibitem[{\citenamefont{Panati et~al.}(2003)\citenamefont{Panati, Teufel, and
  Spohn}}]{Panati}
\bibinfo{author}{\bibfnamefont{G.}~\bibnamefont{Panati}},
  \bibinfo{author}{\bibfnamefont{S.}~\bibnamefont{Teufel}}, \bibnamefont{and}
  \bibinfo{author}{\bibfnamefont{H.}~\bibnamefont{Spohn}},
  \bibinfo{journal}{Commun. Math. Phys.} \textbf{\bibinfo{volume}{242}},
  \bibinfo{pages}{547} (\bibinfo{year}{2003}).

\bibitem[{\citenamefont{Teufel and Spohn}(2002)}]{Teufel}
\bibinfo{author}{\bibfnamefont{S.}~\bibnamefont{Teufel}} \bibnamefont{and}
  \bibinfo{author}{\bibfnamefont{H.}~\bibnamefont{Spohn}},
  \bibinfo{journal}{Rev. Math. Phys.} \textbf{\bibinfo{volume}{14}},
  \bibinfo{pages}{1} (\bibinfo{year}{2002}).

\bibitem[{\citenamefont{Sundaram and Niu}(1999)}]{Sundaram}
\bibinfo{author}{\bibfnamefont{G.}~\bibnamefont{Sundaram}} \bibnamefont{and}
  \bibinfo{author}{\bibfnamefont{Q.}~\bibnamefont{Niu}},
  \bibinfo{journal}{Phys. Rev. B} \textbf{\bibinfo{volume}{59}},
  \bibinfo{pages}{14915} (\bibinfo{year}{1999}).

\bibitem[{\citenamefont{Berry}(1984)}]{Berry}
\bibinfo{author}{\bibfnamefont{M.~V.} \bibnamefont{Berry}},
  \bibinfo{journal}{Proc. R. Soc. A} \textbf{\bibinfo{volume}{392}},
  \bibinfo{pages}{45} (\bibinfo{year}{1984}).

\bibitem[{\citenamefont{Xiao et~al.}(2010)\citenamefont{Xiao, Chang, and
  Niu}}]{Niu_Berry_review}
\bibinfo{author}{\bibfnamefont{D.}~\bibnamefont{Xiao}},
  \bibinfo{author}{\bibfnamefont{M.-C.} \bibnamefont{Chang}}, \bibnamefont{and}
  \bibinfo{author}{\bibfnamefont{Q.}~\bibnamefont{Niu}}, \bibinfo{journal}{Rev.
  Mod. Phys.} \textbf{\bibinfo{volume}{82}}, \bibinfo{pages}{1959}
  (\bibinfo{year}{2010}).

\bibitem[{\citenamefont{Nagaosa et~al.}(2010)\citenamefont{Nagaosa, Sinova,
  Onoda, MacDonald, and Ong}}]{Sinova_AHE_review}
\bibinfo{author}{\bibfnamefont{N.}~\bibnamefont{Nagaosa}},
  \bibinfo{author}{\bibfnamefont{J.}~\bibnamefont{Sinova}},
  \bibinfo{author}{\bibfnamefont{S.}~\bibnamefont{Onoda}},
  \bibinfo{author}{\bibfnamefont{A.~H.} \bibnamefont{MacDonald}},
  \bibnamefont{and} \bibinfo{author}{\bibfnamefont{N.~P.} \bibnamefont{Ong}},
  \bibinfo{journal}{Rev. Mod. Phys.} \textbf{\bibinfo{volume}{82}},
  \bibinfo{pages}{1539} (\bibinfo{year}{2010}).

\bibitem[{\citenamefont{Moore and Orenstein}(2010)}]{Moore_CPGE}
\bibinfo{author}{\bibfnamefont{J.}~\bibnamefont{Moore}} \bibnamefont{and}
  \bibinfo{author}{\bibfnamefont{J.}~\bibnamefont{Orenstein}},
  \bibinfo{journal}{Phys. Rev. Lett.} \textbf{\bibinfo{volume}{105}},
  \bibinfo{pages}{026805} (\bibinfo{year}{2010}).

\bibitem[{\citenamefont{Diehl et~al.}(2007)}]{CPGE_review}
\bibinfo{author}{\bibfnamefont{H.}~\bibnamefont{Diehl}} \bibnamefont{et~al.},
  \bibinfo{journal}{New J. Phys.} \textbf{\bibinfo{volume}{9}},
  \bibinfo{pages}{349} (\bibinfo{year}{2007}).

\bibitem[{\citenamefont{Mineev and Yoshioka}(2010)}]{mineev}
\bibinfo{author}{\bibfnamefont{V.~P.} \bibnamefont{Mineev}} \bibnamefont{and}
  \bibinfo{author}{\bibfnamefont{Y.}~\bibnamefont{Yoshioka}},
  \bibinfo{journal}{Phys. Rev. B} \textbf{\bibinfo{volume}{81}},
  \bibinfo{pages}{094525} (\bibinfo{year}{2010}).

\bibitem[{\citenamefont{Bychkov and Rashba}(1984)}]{Rashba}
\bibinfo{author}{\bibfnamefont{Y.~A.} \bibnamefont{Bychkov}} \bibnamefont{and}
  \bibinfo{author}{\bibfnamefont{E.~I.} \bibnamefont{Rashba}},
  \bibinfo{journal}{J. Phys. C} \textbf{\bibinfo{volume}{17}},
  \bibinfo{pages}{6039} (\bibinfo{year}{1984}).

\bibitem[{\citenamefont{Dugaev et~al.}(2005)\citenamefont{Dugaev, Bruno,
  Taillefumier, Canals, and Lacroix}}]{Dugaev_AHE}
\bibinfo{author}{\bibfnamefont{V.~K.} \bibnamefont{Dugaev}},
  \bibinfo{author}{\bibfnamefont{P.}~\bibnamefont{Bruno}},
  \bibinfo{author}{\bibfnamefont{M.}~\bibnamefont{Taillefumier}},
  \bibinfo{author}{\bibfnamefont{B.}~\bibnamefont{Canals}}, \bibnamefont{and}
  \bibinfo{author}{\bibfnamefont{C.}~\bibnamefont{Lacroix}},
  \bibinfo{journal}{Phys. Rev. B} \textbf{\bibinfo{volume}{71}},
  \bibinfo{pages}{224423} (\bibinfo{year}{2005}).

\bibitem[{\citenamefont{Culcer et~al.}(2003)\citenamefont{Culcer, MacDonald,
  and Niu}}]{Culcer_AHE}
\bibinfo{author}{\bibfnamefont{D.}~\bibnamefont{Culcer}},
  \bibinfo{author}{\bibfnamefont{A.}~\bibnamefont{MacDonald}},
  \bibnamefont{and} \bibinfo{author}{\bibfnamefont{Q.}~\bibnamefont{Niu}},
  \bibinfo{journal}{Phys. Rev. B} \textbf{\bibinfo{volume}{68}},
  \bibinfo{pages}{045327} (\bibinfo{year}{2003}).

\bibitem[{\citenamefont{Avron et~al.}(1983)\citenamefont{Avron, Seiler, and
  Simon}}]{avronseilersimon}
\bibinfo{author}{\bibfnamefont{J.~E.} \bibnamefont{Avron}},
  \bibinfo{author}{\bibfnamefont{R.}~\bibnamefont{Seiler}}, \bibnamefont{and}
  \bibinfo{author}{\bibfnamefont{B.}~\bibnamefont{Simon}},
  \bibinfo{journal}{Phys. Rev. Lett.} \textbf{\bibinfo{volume}{51}},
  \bibinfo{pages}{51} (\bibinfo{year}{1983}).

\bibitem[{\citenamefont{Homes et~al.}(2006)}]{Homes}
\bibinfo{author}{\bibfnamefont{C.~C.} \bibnamefont{Homes}}
  \bibnamefont{et~al.}, \bibinfo{journal}{Phys. Rev. Lett.}
  \textbf{\bibinfo{volume}{96}}, \bibinfo{pages}{257002}
  (\bibinfo{year}{2006}).

\bibitem[{\citenamefont{Adachi et~al.}(2011)\citenamefont{Adachi, Kitajima, and
  Koike}}]{Hall}
\bibinfo{author}{\bibfnamefont{T.}~\bibnamefont{Adachi}},
  \bibinfo{author}{\bibfnamefont{N.}~\bibnamefont{Kitajima}}, \bibnamefont{and}
  \bibinfo{author}{\bibfnamefont{Y.}~\bibnamefont{Koike}},
  \bibinfo{journal}{Phys. Rev. B} \textbf{\bibinfo{volume}{83}},
  \bibinfo{pages}{060506} (\bibinfo{year}{2011}).

\end{thebibliography}
\end{document}